\documentclass[conference]{IEEEtran}
\IEEEoverridecommandlockouts
\usepackage{cite}
\usepackage{amsmath,amssymb,amsfonts}
\usepackage[algo2e]{algorithm2e}
\usepackage{graphicx}
\usepackage{textcomp}
\usepackage{xcolor}
\usepackage{amsmath}
\usepackage{caption}
\usepackage{subcaption}
\usepackage{enumitem}

\SetKwInput{KwInput}{Input}                
\SetKwInput{KwOutput}{Output} 
\usepackage[utf8]{inputenc}
\usepackage[english]{babel}

\usepackage{algorithm,algorithmic,amsmath}
\usepackage{mathtools}
\usepackage{amsmath}
\usepackage{comment}
\def\BibTeX{{\rm B\kern-.05em{\sc i\kern-.025em b}\kern-.08em
    T\kern-.1667em\lower.7ex\hbox{E}\kern-.125emX}}
\usepackage{optidef}

\begin{document}

\title{Latent Neural ODE for Integrating Multi-timescale measurements in Smart Distribution Grids\\
}

\author{
Shweta Dahale, \IEEEmembership{Graduate Student Member,~IEEE}, Sai Munikoti, \IEEEmembership{Graduate Student Member,~IEEE}, 
\\ Balasubramaniam Natarajan, \IEEEmembership{Senior Member,~IEEE} and Rui Yang, \IEEEmembership{ Member,~IEEE} \\

\thanks{S. Dahale is with the Electrical and Computer Engineering, Kansas State University, Manhattan, KS-66506, USA (e-mail: sddahale@ksu.edu). This work was done while she was a summer intern at the National
Renewable Energy Laboratory (NREL) in 2022 summer. Rui Yang is with the Sensing and Predictive Analytics Group in the Power Systems Engineering Center of the National
Renewable Energy Laboratory (NREL), Golden, CO, 80401. S.Munikoti and B. Natarajan are with Electrical and Computer Engineering, Kansas State University, Manhattan, KS-66506, USA, (e-mail:
saimunikoti@ksu.edu, bala@ksu.edu). 

This material is based upon work partly supported by the Department  of  Energy,  Office  of  Energy  Efficiency  and  Renewable Energy  (EERE),  Solar  Energy  Technologies  Office,  under Award Number DE-EE0008767 and the National Renewable Energy Laboratory, operated by Alliance for Sustainable Energy, LLC, for the U.S. Department of Energy (DOE) under Contract No. DE-AC36-08GO28308. Funding provided by U.S. Department of Energy Office of Energy Efficiency and Renewable Energy Solar Energy Technologies Office. The views expressed herein do not necessarily represent the views of the DOE or the U.S. Government. The U.S. Government retains and the publisher, by accepting the article for publication, acknowledges that the U.S. Government retains a nonexclusive, paid-up, irrevocable, worldwide license to publish or reproduce the published form of this work, or allow others to do so, for U.S. Government purposes. 

This work has been accepted in 2023 IEEE ISGT North America conference.
Copyright may be transferred without notice, after which this version may no longer be accessible
}}

\maketitle

\begin{abstract}
Under a smart grid paradigm, there has been an increase in sensor installations to enhance situational awareness. The measurements from these sensors can be leveraged for real-time monitoring, control, and protection. However, these measurements are typically irregularly sampled. These measurements may also be intermittent due to communication bandwidth limitations. To tackle this problem, this paper proposes a novel latent neural ordinary differential equations (LODE) approach to aggregate the unevenly sampled multivariate time-series measurements. The proposed approach is flexible in performing both imputations and predictions while being computationally efficient.
Simulation results on IEEE 37 bus test systems illustrate the efficiency of the proposed
approach.
\end{abstract}


\section{Introduction}

In conventional distribution systems, the lack of sufficient real-time measurements in a distribution grid hinders the ability to obtain situational awareness. With increasing penetration of PV and EVs, more extensive real-time monitoring and control is required for effective operation of the system and for good quality of service to the customers \cite{baran1994state}. To ensure situational awareness, classic distribution system state estimation (DSSE) uses pseudo-measurements with conventional weighted least squares approaches \cite{manitsas2012distribution}. Recently, sparsity-aware DSSE techniques \cite{9247106, dahale2021joint, rout2022dynamic} are proposed to deal with the issue of low observability. 
Alongside, the utilities are upgrading their distribution system to cope with the low observability issue. This has led to a significant increase in the availability of real-time measurements to the control center. For instance, the installation of smart meters and supervisory control and data acquisition (SCADA) sensors have increased to improve the measurement redundancy. Furthermore, new generations of phasor measurement units (PMU) and  IEDs (Intelligent Electronic Devices) enhance situational awareness, protection and control functions in substations. However, aggregating the measurements from different sources presents some challenges. Firstly, the measurements are unevenly sampled i.e., they are obtained at different rates. For example, the smart meter measurements are typically sampled at 15-min interval while the SCADA sensors are sampled at 1-sec to 1-min interval. Furthermore, these measurements may be intermittent or corrupt due to communication network impairments. It is therefore important to reconcile the multi time-scale measurements for situational awareness in a power distribution grid. 

\subsection{Related work}
Research efforts have focused on aggregating two time-scale measurements using linear interpolation/extrapolation based weighted least squares (WLS) approach \cite{gomez2014state}. However, this approach fails to exploit the spatio-temporal relationships in the time-series data and performs poorly in the case of intermittent measurements. The issue of irregular sensor sampling and random communication delays was proposed in  \cite{stankovic2017hybrid}. A multi-task Gaussian process (GP) framework to reconcile heterogeneous measurements was proposed in \cite{9637824}, \cite{dahale2022bayesian}. This approach is computationally expensive as it requires an inversion operation of the kernel matrix to impute or predict the slow-rate measurements. An exponential moving average method to extrapolate the slow-rate measurements was proposed in \cite{karimipour2015extended}.
Authors in \cite{alcaide2017electric} use PMU and SCADA measurements for state estimation. This approach performs DSSE by incorporating a subset of PMU measurements available at time $t$ along with the predicted SCADA measurements. The future predictions of the SCADA measurements are obtained using the information from the previous state estimates. It suffers from large measurement redundancy requirements (around 1.7), which makes it impractical for low-observable distribution systems. Furthermore, \cite{alcaide2017electric} fails to take into account any missing measurements scenario that could occur while aggregating measurements over finite bandwidth communication networks.  A recursive Gaussian process based framework is proposed in \cite{9878081} that sequentially aggregates measurements batch-wise or real-time. The proposed approach requires careful selection of the hyper-parameters of the GP function as well as inversion of the kernel matrix at the initial time step. 

In this paper, we propose a novel approach that leverages spatio-temporal dependencies in time-series data without involving matrix inverse operations. The proposed approach uses neural ordinary differential equations \cite{chen2018neural} that are ideal for imputing and predicting time-series measurements collected at non-uniform intervals. 

\subsection{Contributions}
The main contributions of this paper are summarized below:

\begin{itemize}
    \item For the first time, we propose a latent neural ordinary differential equations (LODE) approach to reconcile the multi time-scale measurements in a distribution grid. The proposed approach is capable of performing imputations as well as predictions at any desired time instant.
    \item The proposed approach is computationally efficient and performs fast predictions once the model is trained offline. 
    \item Simulation results for three-phase unbalanced IEEE 37 bus system reveal the superior performance of the proposed approach. The proposed approach provides smooth imputations and predictions with high fidelity. 
\end{itemize}

\section{Background: Neural ODE}
The measurements obtained from multiple grid sensors are obtained at different sampling rates. The multivariate time-series sensor data with $D$ variables and of $N$ length can be written as,
\begin{equation}
     X_t = x_1, x_2,...,x_T^D \in \mathbb{R}^{N \times D}
     \label{eq:1}
\end{equation}
 
This data may contain missing values due to the sensor sampling rate or communication impairments. A mask $M \in \mathbb{R}^{N \times D}$ identifies the missing measurements. The entries in $M$ i.e., $m_t^d$ is set to 1 if the corresponding measurement $x_t^d$  is observed; else, they are set to 0.
The goal is to reconcile the unevenly sampled  measurements at the finest time resolution. Learning a generative model for the multivariate time-series will help   accomplish this goal.

Generative models using a deep neural network is typically built on the concept of a fixed number of  \textit{layers}. In the forward pass, each network consists of a stack of $L$ transformations, where $L$ is the depth of the model. In order to update these models, a backpropagation algorithm is run through the same $L$ layers via chain rule. This process necessitates that we store the intermediate values of the layers. Thus, training standard deep neural networks are computationally challenging as the memory requirement for storing the intermediate quantity increases as the model depth is increased. Furthermore, limited number of transformations are performed due to the fixed number of layers.   

Neural ODE is a recent novel framework that is effective for modeling irregularly-sampled time series commonly encountered in various real-world application, including smart grid and medical data. It combines deep neural network principles with ordinary differential equations, and thus are more effective than conventional time series models. Particularly, in this work, Neural ODE is used for learning generative models for multivariate time series data from the distribution grid. 
Neural ODE offers a continuous time transformation of variables from input state to final predictions unlike standard deep neural network which only performs a limited number of transformations depending on the number of layers. The transformed values (or intermediate values) are obtained via ODE solvers by providing initial state and dynamics as inputs. The dynamics of the transformation function is determined by a neural network as shown below:
\begin{equation}
    \frac{dz_t}{dt} = f(z_t, \theta)
    \label{eq:2}
\end{equation}
where, $f$ is a neural network parameterized by  $\theta$ that defines the ODE dynamics. $z_t$ is the hidden state of the Neural ODE. Thus, starting from an initial point $z(t_0)$, the transformed state at any time $t_i$ is given by integrating an ODE forward in time, given as,
\begin{equation}
\begin{aligned}
z_i = z_0 + \int_{t_0}^{t_i} \frac{dz_t}{dt} dt \\
z_i = ODESolve (f, z_0, t_0, t_i, \theta)
\end{aligned}
    \label{eq:3}
\end{equation}
\eqref{eq:3} can be solved numerically using any ODE Solver (e.g., Euler's method). In order to train the parameters of the ODE function $f$, an adjoint sensitivity approach is proposed in \cite{chen2018neural}. This approach computes the derivatives of the loss function with respect to the model parameters $\theta$ by solving a second augmented ODE backwards in time. 
Some of the advantages of using Neural ODE solvers over other conventional approaches are: (1) \textit{Memory efficiency:} The adjoint sensitivity approach allows us to train the model with constant memory cost independent of the layers in the ODE function $f$;
(2) \textit{Adaptive computation:} In deep neural networks, the number of layers are fixed and therefore, they generally have a fixed amount of function evaluations. However, the number of layers in a neural ODE is the number of steps an adaptive ODE solver decides to take. This means that the neural ODE can effectively adapt the number of layers on the fly for different datasets and take adaptive steps wherever necessary to determine the solution with desired accuracy; (3) \textit{Effective formulation:} In addition to above two advantages, the continuously defined dynamics can naturally incorporate data which arrives at arbitrary times. Therefore, we propose to use the Neural ODE for reconciling the unevenly sampled distribution grid data. 
\section{Proposed Approach}
\begin{figure*}[h!]
\centering
\includegraphics[width= 0.85\textwidth]{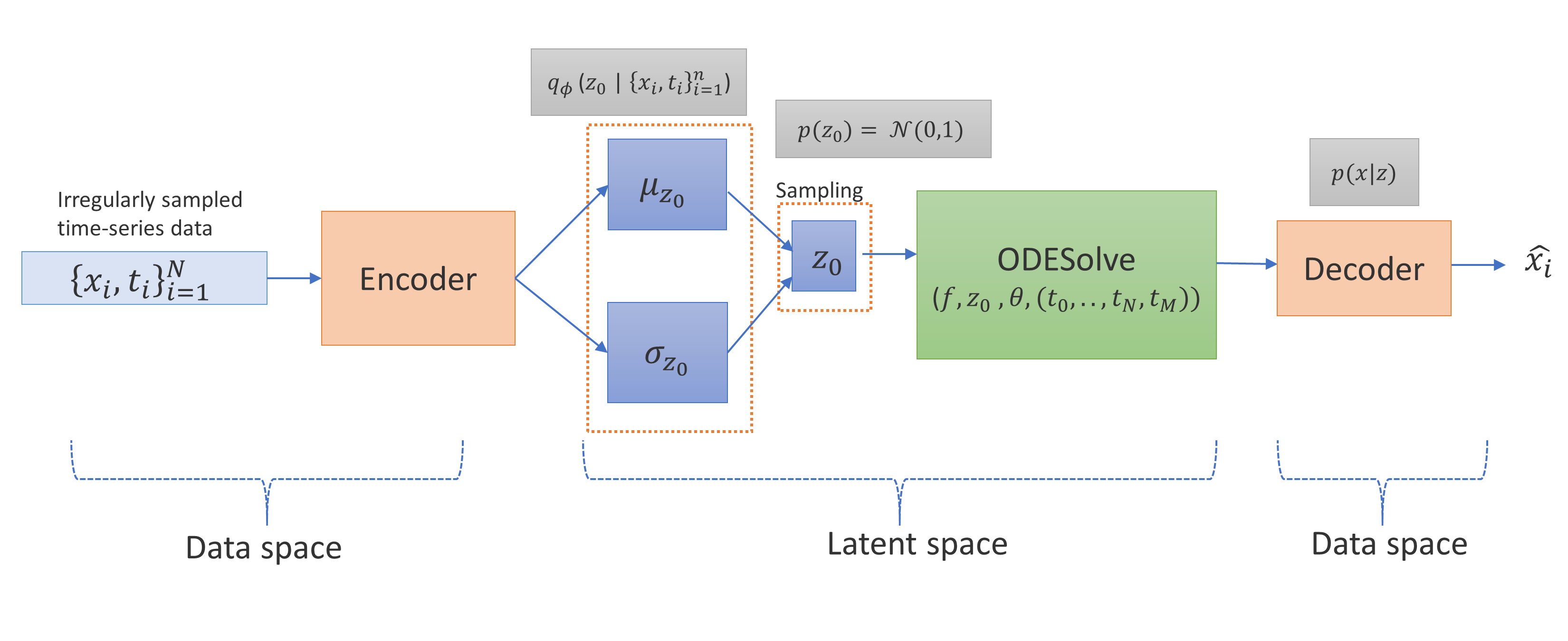}
   \caption{Architecture of proposed Latent ODE approach for smart distribution grid }
     \label{fig:architecture}
 \end{figure*}

The basic neural ODEs evaluate the hidden state values at any desired time instants. However, such models are hard to interpret, especially for multi-time scale power measurements, due to the combined dynamics of power system and the ODE solver. Therefore, we propose to use Latent ODE (LODE) approach for reconciling the multi-time scale power measurements. The LODE approach is a continuous-time generative process for integrating multi-time scale measurements. This approach uses Neural ODEs, and variational autoencoder within a single framework \cite{rubanova2019latent}.
LODE has two key advantages over neural ODE: First, it explicitly decouples the dynamics
of the power system, the likelihood of observations, and the recognition model so that each component can be analyzed separately. Secondly, the posterior distribution over an initial latent state provides a measure of uncertainty which further increases the reliability of our predictions. 
%
The proposed framework (LODE) has three different modules, namely an encoder, a decoder, and the ODE solver. The architecture of LODE for smart distribution grid is illustrated in Fig. \ref{fig:architecture}.
Each module of LODE is described in the forthcoming subsection.

\subsection{Encoder: Recognition network}
This module encodes the input measurements from data space and transforms it into a latent space. Encoding is typically carried out via a recurrent neural network since it is effective in capturing long term dependencies of time series data.
Encoder takes $\{ x_i, t_i\}_{i=1}^{N}$ as an input, where $x_i$ represents the observations and $t_i$ represents the corresponding observation times. The data is processed backward in time from time $t_N$ to $t_0$. An approximate posterior over the initial state  $q_\phi(z_0|\{ x_i, t_i\}_{i=1}^{N})$ is computed from the last hidden layer of the encoder network. In our approach, the mean and standard deviation of the approximate posterior
$q_\phi(z_0|\{ x_i, t_i\}_{i=1}^{N})$ are the function of the final hidden state of an encoder network, characterized by,
\begin{equation}
    q_\phi(z_0|\{ x_i, t_i\}_{i=1}^{N}) = \mathcal{N}(\mu_{z_0}, \sigma_{z_0})  
\end{equation}
where, $\mu_{z_0}, \sigma_{z_0} = g(RNN_{\phi}(\{ x_i, t_i\}_{i=1}^{N}))$. Here, the function $g$ represents a neural network layer, translating the final hidden state of the  encoder into the mean and variance of $z_0$.

\subsection{ ODE Solver}

Once an approximate posterior distribution  $q(z_0|\{ x_i, t_i\}_{i=1}^{N})$ is obtained from the encoder, an initial latent state ($z_0$) for the ODE solver is sampled from the corresponding distribution. The initial latent state serves as an input to the ODE solver together with a ODE dynamic function $f$. Then, a ODE solver is used to obtain latent space observations for all the given times,
\begin{equation}
    z_1, z_2, ..., z_N = ODESolve(f, z_0, \theta,  \{ t_0, t_1,..., t_N\})
    \label{eq:5}
\end{equation}

Thus, given observation times $t_0, t_1, ... , t_N$
and an initial state $z_0$, an ODE solver produces $ z_1, ..., z_N$, which describes the latent state at each observation. The ODE solver is capable of imputing historical time points as well as forecasting future values by providing appropriate latent states at desired time instants. 
As the function $f$ is time-invariant, a unique latent trajectory can be defined given the initial latent state.

\subsection{ Decoder}
the decoder transforms the latent trajectory defined at various time instants back into the data space using the neural network. Standard multi layer perceptron can be used in decoder network since we only need to map two function spaces.

\subsection{Training of LODE}
The training of the proposed LODE framework is similar to that of a variational autoencoder, and it is end-to-end. The encoder-decoder model is trained by maximizing the Evidence Lower Bound (ELBO) given as,
\begin{equation}
    \begin{aligned}
     ELBO(\phi, \theta) = \mathop{\mathbb{E}}_{q_{\phi} (z_0| \{x_i, t_i \}_{i=0}^{N})} [log(p_{\theta} (x_0, ...,x_N))]  \\
         - KL(q_{\phi} (z_0| \{x_i, t_i \}_{i=0}^{N}) || p(z_0))
    \end{aligned}
\end{equation}
The first term in the ELBO represents the log probability of the decoder estimates. The second term denotes the KL divergence or degree of ``dissimilarilty" between the two distributions $q_{\phi} (z_0| \{x_i, t_i \}_{i=0}^{N}) $ and $p(z_0)$. Here, the prior over latent states $p(z_0)$ is chosen as $\mathcal{N}(0,1)$. 
The overall pipeline of the proposed Latent ODE approach summarized in Algorithm 1.
\begin{algorithm}
\KwInput{Datapoints $\{ x_i\}_{i=1}^{N}$ and the corresponding times $\{ t_i\}_{i=1}^{N}$ \\}
\begin{algorithmic}[1]
 
\STATE ${z_0} = RNN(\{ x_i\}_{i=1}^{N})$ 

\STATE $\mu_{z_0}, \sigma_{z_0} = g({z_0})$

\STATE $z_{0} = N(\mu_{z_0}, \sigma_{z_0}) = q_{\phi} (z_0| \{x_i, t_i \}_{i=0}^{N})$

\STATE $z_{1},  z_{2},...,  z_{N} = ODESolve(f, \theta, z_{0}, (t_0,...,t_N))$

\STATE $\hat{x}_i = OutputNN\{ z_{i} \}$
    \STATE \textbf{return} $\hat{x}_i$
  \end{algorithmic}
  \caption{Latent ODE Approach}
\end{algorithm}

\section{Simulation results}
This section evaluates the efficacy of the proposed framework for the imputation and predictions tasks related to smart distribution grid. Experiments are carried out in standard IEEE 37 bus system.

\subsection{Data processing}
We consider measurements from smart meter and SCADA sensors. 
The smart meter measurements consist of 24-hr active and reactive power injection time-series data aggregated at the primary nodes. This 24-hr load profile consists of a mixture of load profiles, i.e., industrial/commercial load profiles \cite{carmona2013fast}, and residential loads \cite{al2016state}. Reactive power profiles are obtained by assuming a power factor of $0.9$ lagging. The SCADA measurements are obtained by executing load flows on the test network. The SCADA measurements consists of the voltage magnitude measurements at a subset of node locations. The aggregated smart meter data are averaged over 15-min intervals while the SCADA measurements are sampled at a 1-min interval. A Gaussian noise with $0$ mean 
and standard deviation equal to $10\%$ of the actual power values is added in the smart meter data to mimic real-world patterns. The smart meter and SCADA measurements constitute our training dataset. 

Once the distribution grid's measurements are obtained, we represent the dataset as a list of records. Each record represents the information about the time-series data with the format given as,
\textit{record = [measurement type, values, times, mask]}. 
Here, time-series data at each node of the IEEE 37 bus network represents one \textit{record}. The \textit{measurement type} denotes the sensor type, i.e., $P,Q,$ or $V$. \textit{Values} $\in \mathbb{R}^{N \times 1}$ represents the sensor measurements with \textit{times} $\in \mathbb{R}^{N}$ as the corresponding time instants. \textit{Mask} $\in \mathbb{R}^{N \times 1}$ represents the availability of the corresponding measurements. The dataset is further normalized between [0,1] interval. We take the union of all time points across different nodes in the dataset that are irregularly sampled. This is needed to perform batching during training. 


\subsection{Model Specifications}

The encoder is a gated recurrent unit (GRU) \cite{cho2014properties}. We consider 40-dimensional hidden states of encoder with tanh activation functions. The ODE function is a feedforward neural network with three layers and $100$ units on each layer.  The ODE solver is a fifth-order `dopri5' solver. The decoder consists of a feedforward neural network with a single layer. Here, we consider an adaptive learning rate with an initial value of the learning rate set to 0.01. We consider batch size as $10$, and report loss as mean squared error (MSE) and negative ELBO. The model is trained using stochastic gradient descent through Adam optimizer for $200$ iterations. All the experiments are conducted on the system with Intel i9 core processor, 32 GB RAM, 8 GB GPU. Python is explicitly used for coding the entire framework with the support of PyTorch's Torchdiffeq package \cite{chen2018neural}. The results of our experiments are discussed in the following subsections.

\subsection{Imputation}

\begin{figure}[h!]
\centering
\includegraphics[width= 0.55\textwidth]{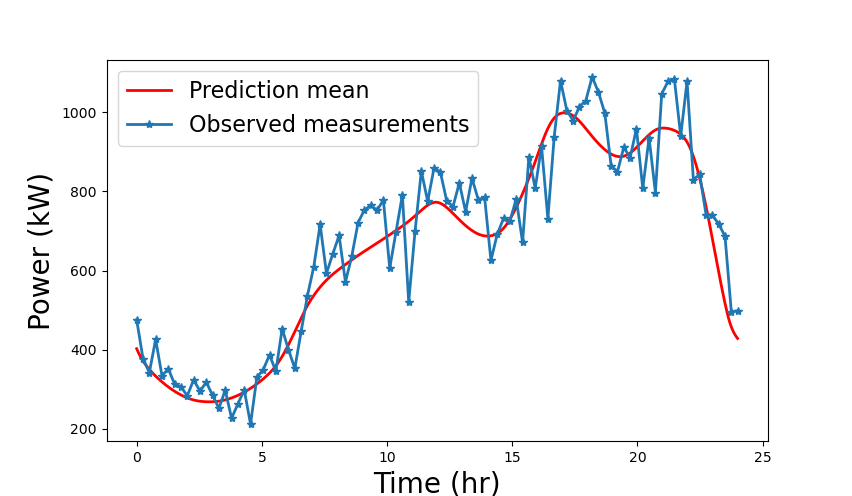}
   \caption{{Imputation at node 1 using Latent ODE approach}}
     \label{fig:ami_node0}
 \end{figure}

\begin{figure}[h!]
\centering
\includegraphics[width= 0.55\textwidth]{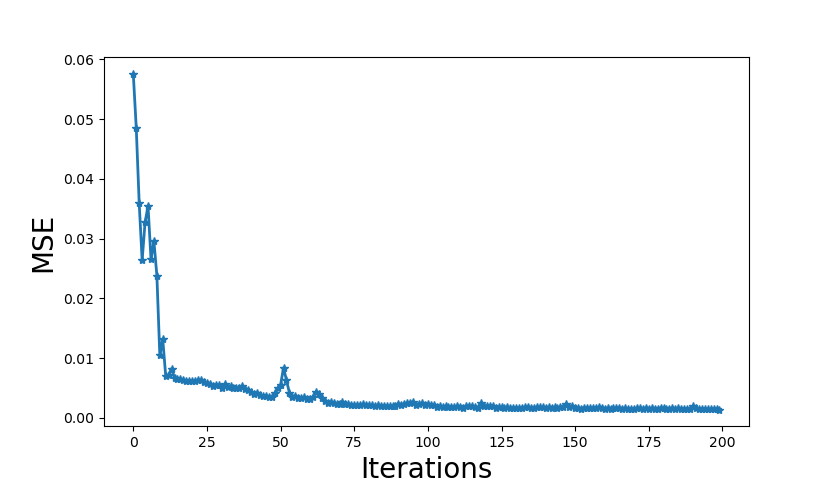}
   \caption{{MSE on the test dataset using Latent ODE approach}}
     \label{fig:testing_loss_mse}
 \end{figure}

\begin{figure}[h!]
\centering
\includegraphics[width= 0.55\textwidth]{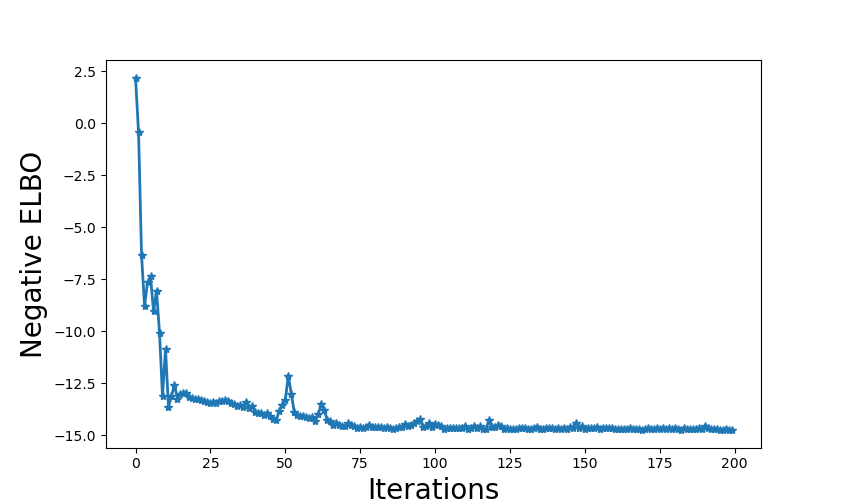}
   \caption{{Negative ELBO on the test dataset using Latent ODE approach}}
     \label{fig:testing_loss_ELBO}
 \end{figure}

In this task, the smart meter measurements are interpolated at a 1-min interval. The training is performed using the observed 15-minute interval data. In order to perform interpolation, the encoder runs backward in time to compute the approximate posterior distribution at the initial time $t_0$. Fig. \ref{fig:ami_node0} demonstrates the imputation performance using the proposed LODE approach. Table. \ref{table1} shows the comparison of LODE approach with linear interpolation approach \cite{gomez2014state} and recursive GP with graphs (RGP-G) approach \cite{9878081}. As seen from Table. \ref{table1}, the proposed approach is accurate with 0.2\% MSE error in the test data. The trajectory of MSE and negative ELBO on the test data are illustrated in Fig. \ref{fig:testing_loss_mse} and Fig. \ref{fig:testing_loss_ELBO}, respectively. The convergence of the losses demonstrate the effective training of the model. 
\begin{table}
\centering
\caption{MSE performance of the proposed LODE, linear interpolation and recursive GP approach}
\label{table1}
\begin{tabular}{|l|l|} 
\hline
\textbf{Approach}                                                             & \textbf{MSE (\%)}  \\ 
\hline
Latent ODE                                                           & 0.2\%     \\ 
\hline
\begin{tabular}[c]{@{}l@{}}Linear \\ Interpolation\end{tabular}      & 2\%       \\ 
\hline
\begin{tabular}[c]{@{}l@{}}Recursive Gaussian~\\process\end{tabular} & 0.7\%     \\
\hline
\end{tabular}
\end{table}

\subsection{Prediction} 

\begin{figure}[h!]
\centering
\includegraphics[width= 0.55\textwidth]{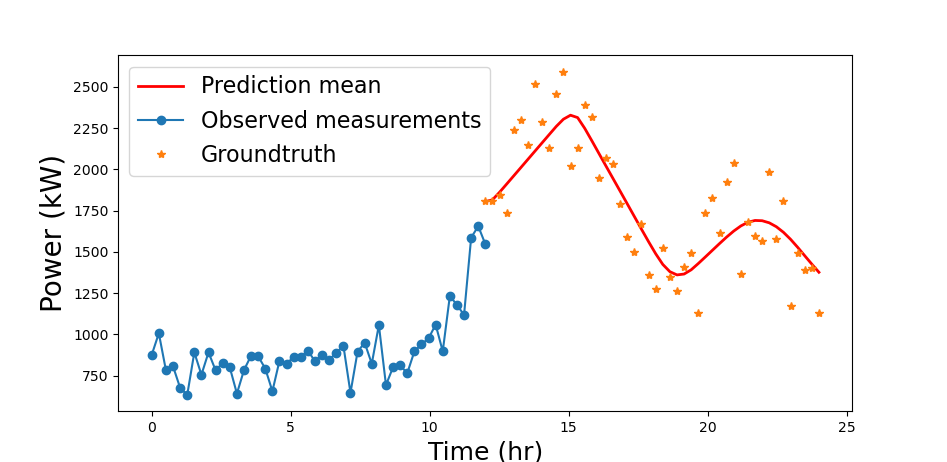}
   \caption{{Predictions at node 12 using Latent ODE approach}}
     \label{fig:extrapolation_node12}
 \end{figure}

In this task, we split the time-series into two halves, $t_0$ to $t_{N/2}$ and $t_{N/2}$ to $t_{N}$. The model is trained by conditioning the observations in the first 12 hrs of the time-series data and reconstructing the other half, i.e., training loss is considered on the second half. Once the model is trained, it can perform predictions for any desired time horizon. As seen from Fig. \ref{fig:extrapolation_node12}, the  model only observes the first 12 hours of measurement data (blue in color) and extrapolates the next 12 hours (red in color). The predictions on the testing data are accurate with MSE of 0.72\%.
 
\section{Conclusion and Future work}
This paper proposes a Latent ODE approach for integrating heterogeneous measurements in a smart distribution grid. The proposed approach uses neural ODEs for learning generative models, which can provide accurate imputations and predictions at any desired time instant. Future work involves developing a robust latent ODE approach against outliers in the measurement data.

\bibliographystyle{IEEEtran} 
\bibliography{ref.bib}
\end{document}